\documentstyle[emulateapj]{article}

\slugcomment{To appear in the PASP (2002 July issue)}

\begin{document}

\title{Possible Recovery of SN 1961V In {\sl Hubble Space Telescope\/} Archival
Images\footnote{Based on observations made with the NASA/ESA {\sl
Hubble Space Telescope}, obtained from the data archive of the Space Telescope
Science Institute, which is operated by the Association of Universities for
Research in Astronomy, Inc., under NASA contract NAS 5-26555.}}

\author{Schuyler D.~Van Dyk} 
\affil{IPAC/Caltech, Mailcode 100-22, Pasadena CA  91125}
\authoremail{vandyk@ipac.caltech.edu}

\author{Alexei V.~Filippenko and Weidong Li}
\affil{Department of Astronomy, 601 Campbell Hall, University of 
California, Berkeley, CA  94720-3411}
\authoremail{alex@astro.berkeley.edu, weidong@astro.berkeley.edu}

\begin{abstract}
SN 1961V in NGC 1058 was originally classified by Fritz Zwicky as a ``Type V''
supernova.  However, it has been argued that SN 1961V was not a genuine
supernova, but instead the superoutburst of an $\eta$ Carinae-like luminous
blue variable star. In particular, Filippenko et al. (1995, AJ, 110, 2261) used
pre-refurbishment {\sl HST\/} WFPC images and the known radio position of SN
1961V to conclude that the star survived the eruption and is likely coincident
with a $V \approx 25.6$ mag, $V-I \approx 1.9$ mag object.  Recently, Stockdale
et al. (2001, AJ, 122, 283) recovered the fading SN 1961V at radio wavelengths
and argue that its behavior is similar that of some Type II supernovae.
We have analyzed post-refurbishment archival {\sl HST\/} WFPC2 data and find
that the new radio position is still consistent with the Filippenko et
al. object, which has not changed in brightness or color, but is also
consistent with an adjacent, fainter ($I \approx 24.3$ mag) and very red ($V-I
> 1.0$ mag) object.  We suggest that this fainter object could be the survivor
of SN 1961V.  Forthcoming {\sl HST\/} observations may settle this issue.
\end{abstract}


\section{Introduction}

Supernova 1961V in NGC 1058 is an unusual object among the nearly 2100
historical supernovae (SNe), and its nature is unclear.  Discovered in 1961
July by P. Wild, SN 1961V had perhaps the most bizarre light curve ever
recorded for a SN, with pre-maximum and post-maximum plateaus of $\sim 0.5$ yr,
a sustained three-year plateau at $m_{\rm pg} \approx 18.5$ mag, and a subsequent
decline to $m_{\rm pg} \approx 21$--22 mag (see Humphreys \& Davidson 1994;
Humphreys, Davidson, \& Smith 1999).  Spectroscopically, it is one of very few
members of the archaic Zwicky (1964, 1965) ``Type V'' SN classification.  Its
spectrum, dominated by narrow emission lines of H, He~I, and Fe~II, suggested a
maximum expansion velocity of only $\sim 2000$ km s$^{-1}$.  Along with only SN
1987A in the Large Magellanic Cloud (LMC; e.g., Gilmozzi et al. 1987;
Sonneborn, Altner, \& Kirshner 1987), SN 1993J in M81 (Aldering, Humphreys, \&
Richmond 1994; Cohen, Darling, \& Porter 1995), and SN 1997bs in M66 (Van Dyk
et al. 2000), SN 1961V has a possible progenitor star identified on pre-SN
images.

The progenitor of SN 1961V may have been one of the most extraordinary stars
known. Identified as a $m_{\rm pg} \approx 18$ mag star in many photographs of NGC
1058 prior to the explosion (Bertola 1964; Zwicky 1964; Klemola 1986), it was
extremely luminous, at $M_{\rm pg}^0 \approx -12$ mag for a distance of 9.3 Mpc
(determined using {\sl Hubble Space Telescope\/} [{\sl HST}] observations of
Cepheids;
Silbermann et al. 1996), leading to an initial (almost certainly erroneous)
theoretical mass estimate of $\sim 2000$ $M_{\sun}$ (Utrobin 1984).
Nonetheless, with somewhat more likely mass estimates of $\gtrsim$100--200
$M_{\sun}$ (Goodrich et al. 1989), it was among the most luminous and massive
known individual stars in any galaxy.

Based on an optical spectroscopic detection of SN 1961V in 1986 and reanalysis
of older data, Goodrich et al. (1989) postulate that SN 1961V was not a genuine
supernova (defined as the explosion of a star at the end of its life), but
rather was the giant eruption of a luminous blue variable (LBV) star, much like
an exaggerated version of $\eta$ Carinae (see Davidson \& Humphreys 1997 for a
review of $\eta$ Car).  Goodrich et al. suggest that SN 1961V actually survived
the eruption, and that the progenitor faded because of the formation of
optically thick dust in the ejecta.  They further predict that at the site of
SN 1961V should be a quiescent S Dor-type, hot Of/WN star enshrouded by dust
and possibly a dense wind; if its circumstellar extinction $A_V \approx 5$ mag
and its bolometric correction BC $\approx -4$ mag, the star should now have $V
\approx 27$ mag.  Alternatively, if $A_V \approx 4$ mag, as in the case of
$\eta$ Car, and BC $\approx -3$ mag, then $V \approx 25$--26 mag.  Finally, $V$
could be as bright as 22--23 mag if the star has $A_V = 4$--5 mag and an
active, optically thick wind with BC $\approx 0$ mag.

Motivated by this prediction, Filippenko et al.~(1995; hereafter F95) attempted
to recover the surviving star via multi-band imaging using 
{\sl HST}, utilizing the radio position obtained by Cowan,
Henry, \& Branch (1988).  Their pre-refurbishment WFPC $VRI$ (flight system
filters F555W, F702W, and F785LP) images, accompanied by precise astrometry
obtained from deep ground-based optical data, led them to associate SN 1961V
with one star in a cluster of three stars, all within $\sim 2\arcsec$ ($\sim
90$ pc) of each other, as the possible eruption survivor.  Their most likely
candidate (``Object 6''), shown in Figure 1 of the present paper\footnote{This
is Figure 2 in F95; it is given again here because it came out so poorly in
F95, both in the original version and in the Erratum.}, has $V \approx 25.6$,
$V-R \approx 1.0$, and $R-I \approx 0.9$ mag, roughly the colors of a
mid-K-type supergiant; if dereddened by reasonable amounts of interstellar and
circumstellar reddening, the colors are consistent with those of O-type stars.
F95 point out that unaberrated images, particularly in bluer bands, are
required to better isolate the possible survivor.

Based on its fading radio emission recently detected in new observations using
the Very Large Array (VLA), Stockdale et al. (2001) argue that the current
radio properties of SN 1961V are consistent with those of some ``peculiar''
Type II radio SNe.  In particular, they compare SN 1961V with SN 1986J in NGC
891 (Weiler, Panagia, \& Sramek 1990), which was originally classified as Type
V (Rupen et al. 1987), but is generally considered to be a prototypical Type IIn
SN (see Schlegel 1990 and Filippenko 1997 for discussion of this SN subtype).
Stockdale et al. contend that it is less likely, based on its radio properties,
that SN 1961V was similar to $\eta$ Car or other LBVs. (However, there is a
lack of radio observations of bright LBVs with ages similar to that of SN
1961V, preventing a secure determination of the true nature of SN 1961V.) In
other words, they argue there should be no survivor, only a very old SN or
young SN remnant emerging forty years after explosion.

In this paper we further investigate whether SN 1961V survived a LBV
superoutburst by exploiting archival {\sl HST\/} WFPC2 images that contain the
SN site.  We recover ``Object 6'' of F95, and we show that an {\it
additional\/} faint and very red object is consistent with the radio position
of SN 1961V and may represent the possible surviving star.  We consider it less
likely to be a fading, very old SN, but additional observations are necessary
to further clarify this issue.

\section{Analysis of Archival {\sl HST\/} WFPC2 Images}

Two {\sl HST\/} programs, GO-5446 and GO-9042, have imaged NGC 1058, the host
galaxy of SN 1961V; the datasets are publicly available in the {\sl HST\/}
archive.  The former program obtained on 1994 September 8 a pair of 80-s
exposures with the F606W filter, while the latter program obtained on 2001
July 3 two pairs of 230-s exposures, one pair with the F450W filter and the
other with the F814W filter. These bandpasses roughly correspond to $V$, $B$,
and $I$.  Using the radio position of SN 1961V from Stockdale et al.~(2001; it
is coincident with the position measured by Cowan et al.~1988, to within the
errors), and correcting for geometric distortions on the WF3 chip, we isolate
the site of the SN for both sets of observations.

Figures 2 and 3 show the SN site on the F814W and F450W images, respectively.
An error circle is plotted at the radio position, based on the astrometric
information in the headers of the F450W and F814W images, with a
(conservative) radius of $\sim 1{\farcs}5$ (the observations by GO-9042
employed the fine-lock mode, which results in about this level of astrometric
uncertainty; see the discussions of {\sl HST\/} astrometry in F95 and Van Dyk
et al. 1999b).  Figure 4 shows the site in the F606W image; this exposure was
obtained in ``gyro'' mode and likely has less reliable astrometry than the
F814W and F450W images, so we impose the error circle of Figures 2 and 3 onto
Figure 4.  The source numbering is from Figure 1, as in F95. Object 8 is
situated off the F450W and F814W images, but is detected on the F606W image.

To measure the brightnesses of the various sources through each of the three
bands, we employed version 1.1 of the package HSTphot (Dolphin 2000a,b).  We
followed the ``recipe'' in the HSTphot
online\footnote{http://www.noao.edu/staff/dolphin/hstphot/} manual, initially adopting a
$4\sigma$ detection threshold and using the following tasks in sequential
order: {\it mask}, {\it crmask}, {\it coadd}, {\it getsky}, {\it hotpixels},
and {\it hstphot}.  As Dolphin (2000b) has shown, HSTphot produces results that
are quite consistent with those obtained from both the DAOPHOT and DoPHOT
packages, while accounting for WFPC2 point-spread function variations and
charge-transfer effects across the chips, zeropoints, aperture corrections,
etc., automatically within one package.  (We have also conducted limited tests
of HSTphot {\it vs.\/} DAOPHOT and find very good agreement in the results.)
Table 1 gives the results of our photometry in the flight system bandpasses.
Not all sources were detected in all three bands, given the relatively low
signal-to-noise ratio (S/N) of these images.  Detection limits ($3\sigma$) are
$m_{\rm F450W} \approx 25.3$, $m_{\rm F606W} \approx 25.4$, and 
$m_{\rm F814W} \approx 25.3$ mag.

In order to further analyze the archival images, we have transformed our flight
system magnitudes into Johnson-Cousins $BVI$ magnitudes, in an analogous manner
to that of F95, i.e., via synthetic photometry of normal stars of a wide span
of spectral types and luminosity classes, obtained by applying the STSDAS
package SYNPHOT to the Bruzual Spectral Synthetic Atlas.  In Table 1 we list
the $BVI$ magnitudes, wherever possible, for the objects in the SN 1961V
environment.  In two cases, we used the detection limit at F606W to set a lower
limit on the $I$ magnitude; for Object 8, we could only approximate the
transformation.

Despite the passband differences between the two studies (F606W and F814W here
{\it vs.\/} F555W and F785LP in F95), the $V$ and $I$ magnitudes of Objects
1--10 agree fairly well overall.  Our $V$-band magnitudes are brighter by
$-0.08$ mag, but with a rather large dispersion, 0.34 mag.  Our $I$-band
magnitudes are fainter by 0.04 mag, with a dispersion of 0.20 mag.  The average
errors in our $V$ and $I$ magnitudes are 0.18 and 0.17 mag, respectively; for
the measurements in F95, they are 0.14 and 0.20 mag, respectively.  Thus, the
$I$-band magnitudes agree to within the errors, but the dispersion in our $V$
magnitudes is about a factor of two larger than the uncertainties in either
study.  We cannot account for this discrepancy entirely through differences in
bandpass or S/N; it may arise from underestimates of the uncertainty in the
aberrated WFPC photometry.  For example, the largest disagreements are for
Object 5 (which is clearly extended), Object 7 (which is in a crowded
environment with a high background), and Object 10 (which also may well be
extended).

In Figures 5 and 6 we show the color-magnitude diagrams for the objects in the
SN 1961V environment.  As pointed out by F95, SN 1961V appears to have occurred
in a region containing several massive supergiant stars.  Objects 4, 7, and 9
are consistent with being early-type supergiants.  Objects 1 and 10 are also of
early type and are possible supergiants, but we cannot be certain because
substantial disagreement exists between their respective $B-V$ and $V-I$
colors.  Object 5 is consistent with either an intermediate-type supergiant or
an early-type supergiant reddened by $A_V \approx 0.8$ mag.  We have no color
information for Object 8, but F95 identified it as an intermediate-type
supergiant.  Both Objects 2 and 6 have quite similar colors and brightnesses,
consistent with late-type supergiants.  An additional faint, red object is
detected only in the F814W image, just to the southeast of Object 6, and was
not identified by F95.  We list this in Table 1 as ``Object 11'' and represent
it with limits in Figure 6 (its $I$-band brightness and $V-I$ color limits are
based on the 3$\sigma$ detection limit of the F606W image).

Figure 3 also hints at an underlying population of fainter blue stars,
extending from Object 5 northeast to Object 8, and to the west around Object
10.  Such a background of blue stars may have also been detected by Utrobin
(1987) and Goodrich et al. (1989).  Figure 4 hints at extended nebular
emission, primarily H$\alpha$ falling within the F606W bandpass, that appears
to trace these blue stars.  The S/N in both the F450W and F606W images is
simply not sufficient to better detect this population and nebulosity.

\section{Discussion}

As can be seen in Figure 2, Objects 2, 5, 6, and 11 are within the positional
error circle for SN 1961V and, therefore, can be considered viable candidates
for its survivor or remnant.  Object 5 is quite diffuse and fairly blue; barely
visible in Figure 2, it is not detected as a source in the F814W image.  F95
identify it as a possible O6 I--III star, although it appears here as a
possible supergiant intermediate in type, or perhaps of early type and
somewhat reddened.  Even if SN 1961V were a fading Type II SN, its color should
be fairly red, even without local interstellar or circumstellar reddening,
since its optical spectral energy distribution should be dominated by H$\alpha$
emission (e.g., Long, Blair, \& Krzeminski 1989; Fesen 1993; Chevalier \&
Fransson 1994).  We therefore do not further consider Object 5 as a possible
candidate.

Both Objects 2 and 6 are quite red, so they are more consistent with our
expectations for either an old SN or reddened LBV star.  They are roughly
equidistant from the radio position (Cowan et al. 1988; Stockdale et al. 2001)
on the WFPC2 images. However, central to the argument by F95 that SN 1961V
survived the superoutburst as Object 6 is its near-coincidence with the radio
position projected onto the WFPC images.  This projection is based on the
application of an independent, accurate astrometric grid, established from deep
ground-based imaging, to the WFPC chip on which the SN environment lies, with
the knowledge that the relative astrometry within a given chip is excellent.
This projection falls within our error circle and may, in fact, better
represent the actual SN position on the {\sl HST\/} images.  Object 2 is more
distant from this position; hence, we no longer consider it to be a candidate.
Furthermore, to within the errors, both Objects 2 and 6 have not changed in
brightness or color between the WFPC and WFPC2 exposures.  It is possible that
they are both LBVs, reddened by similar amounts and remaining at relatively the
same brightness; it is more likely, however, that both Objects 2 and 6 are
similar late-type supergiants.  We therefore also now consider it less likely
that Object 6 is the SN 1961V survivor.

Another candidate for the survivor or very young remnant of SN 1961V may be
Object 11.  It is faint ($I \gtrsim 24.3$ mag), but also quite red ($B-I
\gtrsim 1.0$ mag and $V-I \gtrsim 1.1$ mag).  Unlike the possibly early
stellar-type Object 10, which is actually fainter in $I$, Object 11 is not
detected in the F606W and F450W images.  From the independent astrometric grid,
the radio position actually lies just to the southeast, within a few pixels of
Object 6 (Fig. 1), and is certainly within the WFPC2 error circle.  However, it
is virtually coincident with Object 11.  A hint of this source can be seen in
the deconvolved F785LP image (Fig. 1).  It is seen in each of the raw F814W
images before cosmic-ray rejection and coaddition and is detected by HSTphot in
the coadded image at the 5$\sigma$ level.

Could the brightness and color of Object 11 be consistent with a decades-old
SN?  If SN 1961V is currently an old radio SN, as Stockdale et al.~(2001)
suggest, its optical luminosity, particularly at H$\alpha$, might still be
relatively high, since late-time SN radio and optical emission appear to be
correlated (e.g., Chevalier \& Fransson 1994).  SN 1961V has not dropped off
precipitously in radio brightness at very late times, in contrast with SN 1957D
(Cowan, Roberts, \& Branch 1994; Long, Winkler, \& Blair 1992) or SN 1980K
(Montes et al. 1998), so we also might not expect the optical emission to have
dropped off completely.  Therefore, an old SN 1961V could be bright enough to
detect with {\sl HST\/}, and may be fairly red.

We can explore this further by comparing SN 1961V with two old Type II radio
SNe which are seen in {\sl HST\/} images.  SN 1986J is in publicly available
{\sl HST\/} archival images, obtained by GO-9042 on 2001 July 3 with the F450W
and F814W filters (see Van Dyk et al.~1999b, their Figure 4, for a finder
chart).  Again, applying HSTphot to these images containing SN 1986J, we obtain
$m_{\rm F450W}=23.13{\pm}0.06$ ($B \approx 23.4$ mag) and
$m_{\rm F814W}=20.39{\pm}0.01$ ($I \approx 20.4$ mag), corresponding to a color
$B-I \approx 3.0$ mag.  We can also compare with the Type II-L SN 1979C in
M100, which Van Dyk et al. (1999a) detect in {\sl HST\/} images from 1996.  We
transform the brightness of SN 1979C in the F439W, F555W, and F814W bands to $B
\approx 23.2$ mag, $V \approx 22.1$ mag, and $I \approx 21.0$ mag; hence, $B-I
\approx 2.2$ and $V-I \approx 1.1$ mag.  Both SNe are quite red, but the
brightness of SN 1986J is evolving much more slowly than that of SN 1979C
(compare $I \approx 20.4$ mag for SN 1986J with Gunn $i \approx 20.2$ mag in
1986; Rupen et al. 1987).

The host galaxies for SNe 1961V and 1986J are in the same group, at a distance
of 9.3 Mpc (Silbermann et al. 1996).  If we assume $A_V = 1.5$ mag for SN 1986J
(Leibundgut et al. 1991), we find that $M_I^0 \approx -10.3$ mag.  For SN
1979C, at 16.1 Mpc and $A_V \approx 0.5$ mag (Van Dyk et al. 1999a), $M_I^0
\approx -10.5$ mag.  (If either SN is experiencing higher intrinsic
circumstellar extinction, then these absolute magnitudes will be even
brighter.)  In contrast, for Object 11, assuming circumstellar $A_V \approx
4$--5 mag, we derive $M_I^0 \approx -7.9$ to $-8.5$ mag; assuming less
extinction, of course, means that Object 11 would be intrinsically fainter.

Making a direct comparison of SN 1961V in the optical with these old radio SNe,
of course, is difficult, since both SN 1986J and SN 1979C are about 20 years
younger than SN 1961V \footnote{Although not discovered until 1986, the
explosion date of SN 1986J was possibly sometime between 1978 and 1983 (Rupen
et al.~1987).}; they could still significantly fade over the next two
decades. Also, SNe~II-L and IIn behave differently in both the optical and the
radio.  Unfortunately, no optical data on old radio SNe as old as SN 1961V are
available, other than the ground-based emission-line data for SN 1957D from
1987 through 1988 (Long et al. 1989).  Nonetheless, although SNe 1986J and
1979C are indeed quite red, they both appear to be intrinsically brighter than
Object 11, making it less likely that Object 11 is an old SN.

Another distinct possibility is that no positional coincidence actually exists
--- SN 1961V is not Object 11 or any of the other objects detected in the WFPC2
images. This implies that SN 1961V may have faded below detectability in all of
these bands (specifically, it must have $I \gtrsim 25.3$ mag).  If SN 1961V is
an old SN, then it would have intrinsically faded or circumstellar extinction
would have significantly increased since the 1986 spectroscopic observations by
Goodrich et al. (1989).  (In fact, if this is true, then the SN must also have
faded significantly between 1986 and the 1991 WFPC observations by F95.)  Even
if SN 1961V is an $\eta$ Car-like LBV that has survived a superoutburst, it
also may no longer be detectable: A similar, younger case of a possible $\eta$
Car-like LBV, SN 1997bs in M66 (Van Dyk et al. 2000), has faded considerably
and become bluer in recent {\sl HST\/} Snapshot images (Li et al. 2002), rather
than redder and brighter, as predicted for increasing amounts of dust.

\section{Conclusions}

Ultimately, we conclude that SN 1961V is most likely Object 11, and that it has
survived its eruption.  The brightness and color limits for Object 11, shown in
Figure 6, are consistent with an early-type supergiant star with $A_V \gtrsim
1$ mag --- specifically, the $V$ magnitude ($\sim 27$--25 mag) predicted by
Goodrich et al. (1989) for circumstellar $A_V \approx 4$--5 mag and BC $\approx
-3$ to $-4$ mag of a quiescent LBV surrounded by optically thick dust and a
stellar wind.  Its current absolute brightness is more consistent with a
low-luminosity LBV, such as R71 in the LMC (Humphreys \& Davidson 1994),
possibly in quiescence, than with a decades-old ``peculiar'' Type II SN.

Clearly, full resolution of the nature of SN 1961V still eludes us.  As
Stockdale et al. (2001) point out, further multi-wavelength observations are
needed.  Ideally, one might obtain spectra of the candidates: Gruendl et
al. (2002) attempted to detect the SN with the echelle spectrograph on the Kitt
Peak 4-m telescope, but failed to do so.  They conclude that they simply may
have misplaced the spectrograph slit.  However, if SN 1961V is indeed Object
11, an alternative explanation is that its relative faintness and redness may
have made detection very difficult.

An upcoming {\sl HST\/} Cycle 11 program, GO-9371 (PI: Y.-H. Chu), will obtain
spectra of the SN 1961V environment within $1''$ of the SN position using STIS,
to find signatures of LBV ejecta nebulae, old SNe, and SN remnants.  They will
also obtain WFPC2 $VRI$ images.  If the slit is centered on Object 11, and the
exposures are of sufficient depth, these spectra might provide the definitive
answer.  If the broad-band imaging is also sufficiently deep and at the highest
possible resolution, the brightnesses and colors of the candidate and other
stars in the environment could be more accurately measured.  SN 1961V is also a
GTO target using the newly installed ACS at the F475W, F625W, and F775W bands,
complementing the GO program.

In addition, deep, high-resolution, near-infrared imaging may be necessary,
since we might expect old, cooling SNe and dust-enshrouded hot LBVs to have
different infrared colors.  For example, $\eta$ Car has $J-H \approx 0.9$ mag
and $H-K \approx 1.5$ mag (Whitelock et al. 1983), while at late times
($\gtrsim$100 d) the Type IIn SN 1998S (also a radio SN) has $J-H \approx 0.4$
mag and $H-K \approx 0.8$ mag due to dust in the circumstellar medium (Fassia
et al. 2000).  Unfortunately, published data are not available for any SN~IIn
(and, with the exception of SN 1987A, for any other SN of any type) in the
infrared at very late times (i.e., several years to decades after explosion),
but we might expect that these SNe get redder through additional dust
formation.  Although this is not a rigorous test, the infrared detection of SN
1961V would also help determine its nature.  The red color of Object 11 implies
that it should be a relatively bright infrared source for detection with a
large-aperture telescope under excellent seeing conditions, or with 
{\sl HST}/NICMOS and {\sl SIRTF}.

Finally, as an important aside, if SN 1961V, SN 1986J, the old ``Type V'' SN
classification, and Type IIn SNe bear any connection, {\it and\/} if the
progenitor of SN 1961V did survive, then the ``explosion'' mechanism is not
core collapse in all SNe IIn.  That is, the SN~IIn subclass, which is already
known to span a very broad range of properties (e.g., Filippenko 1997), may
include a number of SN ``impostors.''  The occurrence of SN 1961V and its
possible cousins [SN 1954J (Smith, Humphreys, \& Gehrz 2001), SN 1997bs (Van
Dyk et al. 2000), SN 1999bw (Filippenko, Li, \& Modjaz 1999), and SN 2000ch
(Filippenko 2000)], and their apparent resemblance to $\eta$ Car and LBV
superoutbursts, add an intriguing twist to the evolution of very massive stars.

\acknowledgements

We thank A.~J.~Barth for producing Figure 1, which was originally published by
Filippenko et al.~(1995).  The work of A.V.F.'s group at UC Berkeley is
supported by NSF grant AST-9987438, as well as by NASA grant AR-08754 from the
Space Telescope Science Institute, which is operated by AURA, Inc., under NASA
contract NAS5-26555.  A.V.F.~is grateful for a Guggenheim Foundation
Fellowship.

\begin{deluxetable}{cccclll}
\def\phmm{\phm{$-$}}
\tablenum{1}
\tablecolumns{3}
\tablecaption{{\sl HST\/} WFPC2 Photometry of the SN 1961V Environment}
\tablehead{\colhead{Object} & \colhead{$m_{\rm F450W}$} 
& \colhead{$m_{\rm F606W}$} & \colhead{$m_{\rm F814W}$}
& \colhead{$B$} & \colhead{$V$}
& \colhead{$I$}}
\startdata
1 & 24.51(02) & 25.07(35) & 24.15(24) &   24.49 & 24.97   & 24.15    \nl
2 & \nodata & 24.95(24) & 23.57(12)   & \nodata & 25.47   & 23.58    \nl
3 & \nodata & \nodata & 23.15(08)     & \nodata & \nodata & \llap{$\gtrsim$}23.2 \nl
4 & 23.84(10) & 24.22(14) & 24.09(17) &   23.82 & 24.22   & 24.10    \nl
5 & 24.76(25) & 24.39(16) & \nodata   &   24.82 & 24.50   & \nodata  \nl
6 & \nodata & 25.00(27) & 23.73(13)   & \nodata & 25.49   & 23.73    \nl
7 & 24.02(14) & 23.84(14) & 23.83(14) &   24.04 & 23.85   & 23.83    \nl
8 & \nodata & 24.29(01) & \nodata     & \nodata & \llap{$\sim$}24.3 & \nodata \nl
9 & 23.63(10) & 23.84(10) & 23.76(13) &   23.61 & 23.83   & 23.76    \nl
10 & 24.79(24) & 24.07(12) & 24.48(26) &  24.90 & 24.13 &   24.46    \nl
11 & \nodata & \nodata & 24.31(22)    & \nodata & \nodata & \llap{$\gtrsim$}24.3 \nl
\enddata
\tablenotetext{}{Note: The values given in parentheses are the 
uncertainties in the last two digits of the magnitudes.}
\end{deluxetable}

\clearpage

\begin{figure}
\figurenum{1}

\caption{Figure 2 of F95, showing {\sl HST\/} WFPC unrestored ({\it left}) and
deconvolved ({\it right}; Lucy-Richardson algorithm) images of the SN 1961V
environment.  Objects discussed in the present paper are labeled. The nominal
radio positions (Cowan et al. 1988; uncertainties $\pm 0.3''$) of SN 1961V and
an old radio supernova remnant are marked with triangles labeled ``61V'' and
``SNR,'' respectively.}

\end{figure}

\begin{figure}
\figurenum{2}
\caption{{\sl HST\/} WFPC2 F814W image of the SN 1961V environment, obtained by
program GO-9042 and available in the archive. An error circle of radius
$1.5''$ is plotted at the radio position of SN 1961V.
Note how close the environment
is to the edge of the WF3 chip. Source numbering is
given in Figure 1. Also labelled is Object 11, which is coincident with the
radio position of SN 1961V, as indicated by F95, and which we suggest may be
the survivor of SN 1961V.}
\end{figure}

\begin{figure}
\figurenum{3}
\caption{{\sl HST\/} WFPC2 F450W image of the SN 1961V environment,
obtained by program GO-9042, with the same scale,
position, and orientation as in Figure 2.  Object 8, as in Figure 2, is off 
the WF3 chip.}
\end{figure}

\begin{figure}
\figurenum{4}
\caption{{\sl HST\/} WFPC2 F606W image of the SN 1961V environment,
obtained by program GO-5446, with the same scale,
position, and orientation as Figure 2.  Object 8, which is off the WF3 
chip in Figures 2 and 3, is labelled.}
\end{figure}

\clearpage

\begin{figure}
\figurenum{5}
\plotone{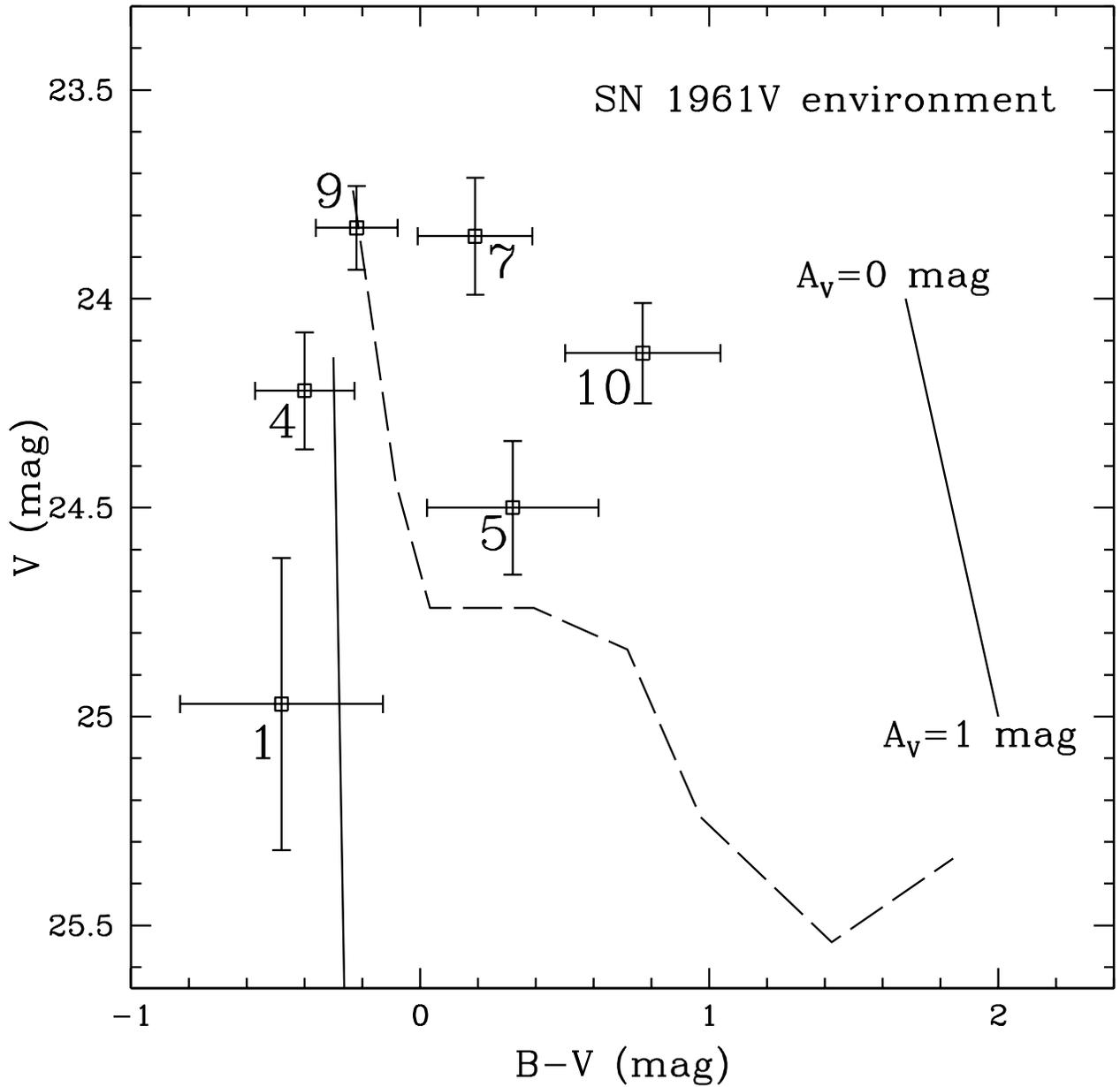}
\caption{$B-V$ {\it vs.\/} $V$ color-magnitude diagram for the objects
in the SN 1961V environment.  The points for the objects are labelled.
Shown is the locus for the main sequence ({\it solid line}) and for 
supergiants ({\it dashed line}) derived from Bessell (1990) and Binney 
\& Merrifield (1998).  Also shown is the reddening vector from Cardelli 
et al.~(1989).}
\end{figure}

\clearpage

\begin{figure}
\figurenum{6}
\plotone{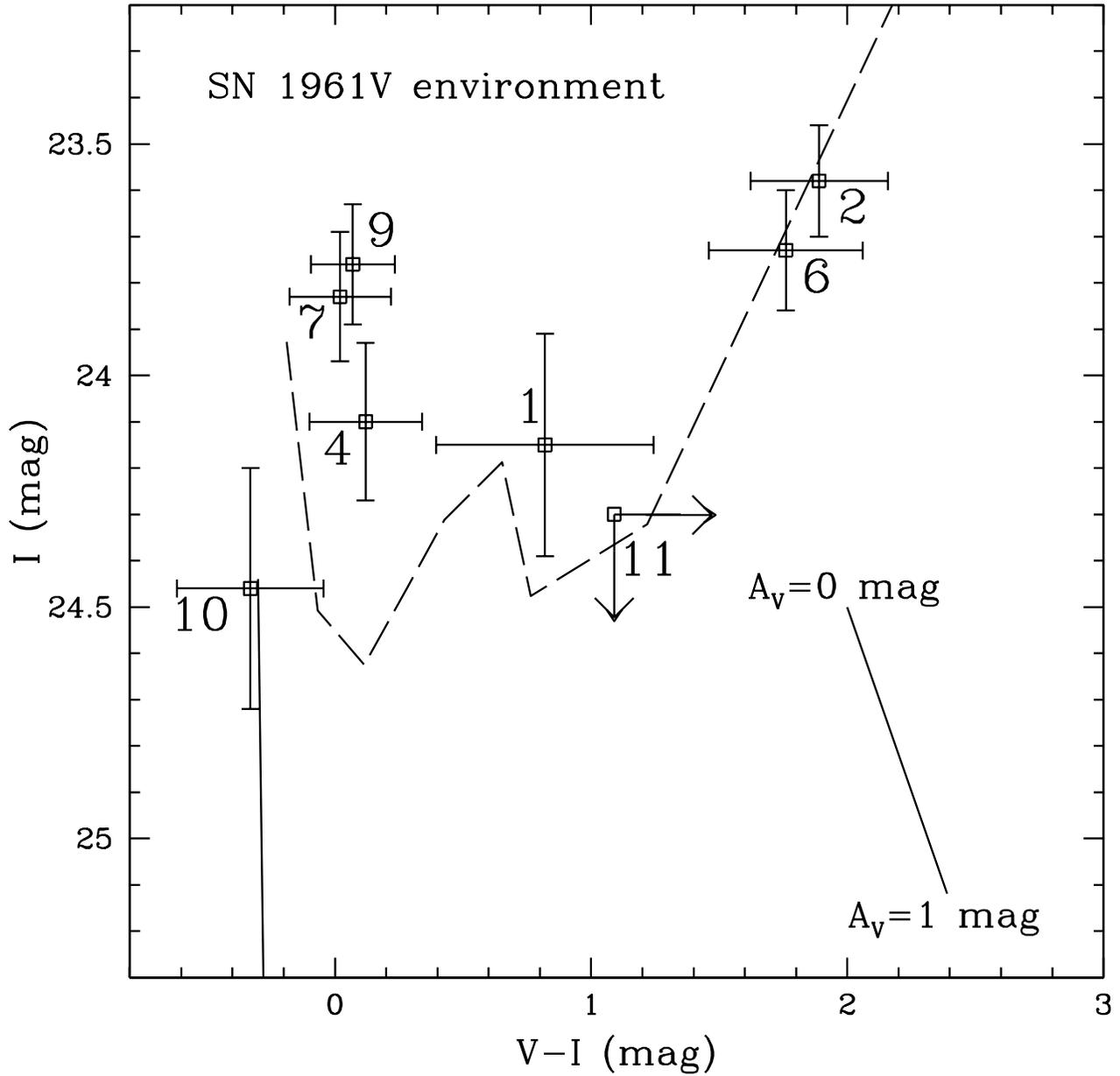}
\caption{$V-I$ {\it vs.\/} $I$ color-magnitude diagram for the objects
in the SN 1961V environment.  The points for the objects are labelled.
An upper limit for the magnitude and lower limit for the color of Object 
11 is represented.  Shown are the loci for the main sequence ({\it solid 
line}) and for supergiants ({\it dashed line}) derived from Bessell (1990) 
and Binney \& Merrifield (1998).  Also shown is the reddening vector from 
Cardelli et al.~(1989).}
\end{figure}

\end{document}